\documentstyle[12pt]{article}
\begin{document}
\baselineskip = 20pt

\begin{flushright}
{IF-UFRJ/Mar-02}
\end{flushright}

\vspace*{5mm}
\begin{center}

{\bf  String Cosmology in the Jackiw-Telteiboim Model with Quantum Corrections   }
\footnote{Work supported in part by 
Funda\c c\~ao Universit\'aria Jos\'e Bonif\'acio, FUJB.}
\vspace*{1cm}

\end{center}
\begin{center}

{M. Alves$^{a}$}
\end{center}

\begin{center}
{Instituto de F\'\i sica \\ Universidade Federal do Rio de Janeiro \\
 Caixa Postal 68528  Rio de Janeiro 21970-970 Brazil}
\end{center}
\vskip 1cm
\vspace*{5mm}
\begin{center}
{\bf ABSTRACT}
\end{center}

\vspace*{2mm}

\noindent

This note deals with the possibility of non-trivial cosmological solutions given by quantum corrections in the framework of the Jackiw-Telteiboim model to the bidimensional gravity. The resulting model shows that the quantum corrections transform, in some cases, the classical solution into a more interesting one with  initial singularity.    

\bigskip

\vfill

\par

\bigskip

\bigskip

\par

\bigskip

\bigskip

\noindent $(a)${msalves@if.ufrj.br}

\pagebreak

{\noindent\bf 1-Introduction}

It is widely recognized that two-dimensional models of gravity can give us a better understanding of the gravitational quantum effects. These models, derived from the string motivated effective action [1] or from the dimensional reduction of the Einstein equations [2] have a rich structure in spite of their relative simplicity. Gravitational collapse, black holes and quantum effects are examples of subjects whose description is rather complicated in four dimensional gravity while their lower-dimensional version turns to be more treatable, sometimes completely solved.

 By two-dimensional gravity we mean a relativistic theory of the metric tensor in two-dimensions (1+1) space-time so, besides the above mentioned interpretations, we can realize some properties of the string theory by considering these models as a theory that describe the evolution of the string in terms of the two parameters, embedded in a higher dimension space-time [3].

 In a series of previous works we show that the black hole solutions are modified when quantum corrections are considered in the JT  and the CGHS models. In the first model, black hole solutions arise from the quantum correction even when we have no mass nor cosmological constant [4] and for the second theory, we have a modification of the expression of the Hawking radiation [5].   

In this note we work out a bidimensional cosmology in the framework of the Jackiw-Teiltelboim theory to gravitation (JT)[1] when quantum effects are considered  via a generalized expression to the trace anomaly (we will clarify these points soon). The scale factors found show nontrivial behavior even in the absence of source of matter with cosmological interpretation and we note that these effects are dominant near the singularity. These are the main results of the work.  

The paper is organized as follows: we present the JT theory and the cosmological model that we will use to derive the solutions. After that, we discuss the trace anomaly and the required generalized expression as well. In the following section, the modified model is presented and solved to a few cases of interest. General comments and remarks are in the last section.      

\bigskip
{\noindent\bf 2-The Jackiw-Teiltelboim Model and Bidimensional Cosmology}

The well known difficulty of the bidimensional gravity due to the absence of dynamics of the Einstein-Cartan action in (1+1) space-time can be circumvent by using the JT model: We start directly from the equation of motion 

\begin{equation}
R- \Lambda =  T , 
\end{equation} 

\noindent where $R$ is the scalar curvature, $\Lambda$ is the 2d version of the cosmological constant and $T=T^{\mu}_{\mu}$  the trace of energy-momentum tensor (TEM) of the matter source.

The solutions that we are interested on is given by the general Robertson-Walker metric line element, 

\begin{equation}
ds^2= -dt^2 + a(t)^2dx^2
\end{equation}

\noindent with a(t) being the scale factor.

A fundamental quantity here is the scalar curvature that, with the metric (2), is 

\begin{equation}
R=2{\ddot a(t)\over a(t)},
\end{equation}
 
\noindent and contains the essential geometric informations in two dimensions.

Just to compare results, lets us solve the simplest two cases of (1) namely , the massless case, with $\Lambda=0$ and $\Lambda\not= 0$. The first one is the flat spacetime with

\begin{equation}
a(t)=At+B
\end{equation}
and a de Sitter type  solution with $\Lambda \not=0$, 

\begin{equation}
a(t)=Ae^{\Lambda t} + Be^{-\Lambda t},
\end{equation}

\noindent with $A$ and $B$ constants that can be adjusted. In this case, to $A=1$ and $B=0$, there is a coincidence with the four dimensional solution, which is the common limiting case for $t\rightarrow 0$ of all the models.

The next step is the introduction of the quantum corrections to investigate the possibility of modifications of the solutions (4) and (5), for instance. The most natural procedure in this direction is to consider the matter source, given for the right side of (1), quantized on a classical background space time, $ <T>=g_{\mu\nu}<T^{\mu\nu}>$. The signal of this contribution is the trace anomaly: the fact that the expected value of the TEM of any massless field is non zero, in contrary to the classical value. However, as mentioned in [4], the renormalized value of the anomaly is just the scalar curvature  $R$,  and we conclude that this will leave (2) essentially the same, modulo a rescaling of the constants. Fortunately this is not the end of the history: as a matter of consistency, the semiclassical theory must to satisfy the so called Wald axioms [7]. The first one is that the expected value of the TEM must be covariantly conserved and it is ensured by using a modified expression for the anomaly which introduces non-trivial contributions. In the next section we will discuss these points in detail. 

\bigskip
\bigskip

{\noindent\bf 3 -A Generalized Expression of the Trace Anomaly}

The modified expression of the 2d trace anomaly is derived in a series of recent papers [6].  The exact expressions are not the same in some of them, but these discrepancies are essentially  in the numerical coefficients and will not affect both results and discussions here.

Using the notation adopted by us in [7], the above mentioned generalization of the anomaly is 

\begin{equation}
<T>=R + \nabla^{\mu} A_{\mu} + A^{\mu}A_{\mu}
\end{equation}
  
\noindent with 

\begin{equation}
A_{\mu} = (-g)^{1/4}\partial_{\mu}(-g)^{(1/4)}= -{1\over 2} {\dot a\over a}
\end{equation}

\noindent using (2).  Some of the new terms in (6) can be reabsorbed in  the expression of $<T>$, resulting

\begin{equation}
<T>=  {\ddot a \over a} + \alpha ({\dot a\over a})^{2},
\end{equation}

\noindent where $\alpha$ is to point us the quantum correction in the model and it is this term that gives us the non-trivial contribution in the JT model:

 \begin{equation}
 R- \Lambda =<T>=  {\ddot a \over a} + \alpha ({\dot a\over a})^{2},
\end{equation}

\noindent or, using (3) and redefining again the constants, 

 \begin{equation}
  {\ddot a \over a} + \alpha ({\dot a\over a})^{2}=\Lambda.
\end{equation}

  This is our basic equation and it will be  valid to all massless matter fields.  In the following we will analyze the solutions of (10) and compare with those found in the classical  cases (4) and (5).  

\bigskip
\bigskip

{\bf 4- The  Solutions with Quantum Corrections}

\noindent i) $\Lambda$ =0

This is the modified version of the trivial case (4) and the solution is easily found to be, from (10),

\begin{equation}
a(t)\,\,=\,\, \left( A(1+\alpha)t + B(1+\alpha )\right) ^{{1\over 1+\alpha}}
\end{equation}

\noindent that is, for small $\alpha$, 

\begin{equation}
a(t)\,\,= (At+B)+\alpha (At+B)(1-ln(At+B)) + O(\alpha^{2}).                   
\end{equation}

The corresponding curvature scalar is, using (3), 

 \begin{equation}
  R=\,\, -2{A^{2} \alpha\over ((At+B)(At+B+(At+B)(1-ln(At+B))\alpha))}             
\end{equation}
\noindent that, to small $\alpha$, results   

\begin{equation}
R\,=\, -2\alpha {A^{2}\over (At+B)^{2}}
\end{equation}

So, the anomaly, indicated by $\alpha$, adds a non-zero value to the curvature scalar. Moreover, the new term has a singularity at the (shifted) initial time $ t^{'} = t + {B\over A}$. Note that  this is the expression for the four-dimensional Einstein-De Sitter model, the behavior of all big bang models, in (1+3), at the early stages  ($R \rightarrow 1/t^2$).

\bigskip
\bigskip

\noindent ii) $\Lambda$ $\not= 0$

The solution with $\Lambda >0$ can cast, without lost of generality, as

\begin{equation}
a(t)=-A(1-\alpha){e^{\left({\sqrt{G(1+\alpha)}t\over 1+b}\right)}\over \sqrt{G(1+\alpha)}}.
\end{equation}

This solution, however, does not introduce any modification to the behavior of the classical version (5): the curvature scalar is given by  

\begin{equation}
R\,=\,  {G \over 1+\alpha}\,\,=\,G^{'}
\end{equation}

\noindent and makes just a rescaling of the cosmological constant.

The same does not occur when we consider the anti-DeSitter case to wit, $\Lambda <0$ or 

 \begin{equation}
  {\ddot a \over a} + \alpha ({\dot a\over a})^{2}=-\Lambda
\end{equation} 

\noindent with a solution as 

 \begin{equation}
  a(t)\, =\, \left({A cos(G\sqrt{1+\alpha}t)\over G}\right)^{{1\over 1+\alpha }}.
\end{equation}

Differently of the DeSitter solution (15), here the quantum correction gives a non trivial term to $R$ , that is, to small $\alpha$ (and $A=1$),

 \begin{equation}
R\,=\,   -\Lambda -\alpha\,\, tan(Gt)^{2}.
\end{equation} 

A few comments is in order here: the quantum corrections does not affect the (2d) de Sitter solution (5) and the $t\rightarrow 0$ behavior remains the same. In the Anti-de Sitter case, the solution remains oscillatory but now there is a divergence in $R$ to some values of $t$. Both solutions have the same limit at early times. The Einstein-de Sitter semiclassical case (12) turns out to be like the four dimensional one since now we have a mass scale arising from the trace anomaly.

\bigskip
\bigskip

{\bf 5- Conclusion}

In this paper , we worked out the semiclassical version to the JT model for 2d gravity for massless field as the matter source. We made use of the improved version of the trace anomaly otherwise the model would be transparent to the quantum nature of the fields.  

The solutions found here show us that the quantum correction implies the possibility of non trivial cosmologies. The solution without cosmological constant and that with negative value are drastically transformed. The solution to the de Sitter case remains the same.   

It is recognized that quantum effects must be relevant in situations with high values of the curvature. An example of this phenomenon is the Hawking-Bekestein radiation of the black-hole. The results derived in this note, in spite of the simplicity of the model, are in the same direction: in eqs (14) and (19) the contribution of the anomaly, signalized by $\alpha$,  turns out to be dominant close to the singularity independently of the value of this parameter. 
It is worth to mention, that these results are derived from a semiclassical theory and must be taken as  approximated, first order, ones since in these regions  a full quantized theory would be necessary.

\pagebreak

{\bf 4- References}

\noindent 1. R.Jackiw, in Quantum Theory of Gravity, ed. S.Christensen (Adam Hilger, Bristol, 1984),p.403; C.Teiltelboim, ibid.,p.327. R.Mann, A.Shiekm and L.Tarasov, Nucl. Phys. {\bf B341},134 (1992).

\noindent 2. C.G.Callan, S.B.Giddings and J.A.Strominger, Phys.Rev.{\bf D45},R1005(1992); J.A.Strominger, in Les Houches Lectures on Black Holes (1994), hep-th/9501071.

\noindent 3. A.Polyakov, Mod.Phys.Lett.A2,89391987); Gauge Fields and Strings (Harwood, New York,NY,1987).

\noindent 4. M.Alves, Class.Quantum Grav. 13,171(1996).

\noindent 5. M.Alves, Int.J.Mod.Phys.D8,6,687(1999). 

\noindent 6. R.M.Wald, Commun. Math. Phys.,54,1(1977);N.D.Birrel and P.C.Davies, in {\it Quantum Fields in Curved Spacetime} (Cambridge University Press, Cambridge,1984) .

\noindent 7. M.Alves and C.Farina, Class. Quantum Grav. 9, 1841 (1992); W.Kummer, H.Liebl and D.V.Vassilevich, Mod.Phys.Lett {\bf A 12 }, 2683 (1997); S.Hawking and R.Boussos, hep-th/9705236; J.S.Dowker, hep-th/9802029; S.Ichinoise and S.Odintsov, hep-th/9802043.

\end{document}